\documentclass[aip, apl, longbibliography, amsmath, amssymb, reprint, noshowpacs, floatfix, superscriptaddress]{revtex4-2}

\usepackage{graphicx}
\usepackage{dcolumn}
\usepackage{bm}

\usepackage[utf8]{inputenc}
\usepackage[T1]{fontenc}
\usepackage{mathptmx}
\usepackage{etoolbox}

\makeatletter
\def\@email#1#2{%
 \endgroup
 \patchcmd{\titleblock@produce}
  {\frontmatter@RRAPformat}
  {\frontmatter@RRAPformat{\produce@RRAP{*#1\href{mailto:#2}{#2}}}\frontmatter@RRAPformat}
  {}{}
}%
\makeatother

\begin{document}
\title{Nanoscale Thermal Imaging of VO$_2$ via Poole-Frenkel Conduction}

\author{Alyson Spitzig}
\affiliation{Department of Physics, Harvard University, Cambridge, Massachusetts 02138 USA}
\affiliation{Department of Physics and Astronomy, University of British Columbia, Vancouver, British Columbia V6T 1Z4 Canada}

\author{Adam Pivonka}
\author{Alex Frenzel}
\author{Jeehoon Kim}
\affiliation{Department of Physics, Harvard University, Cambridge, Massachusetts 02138 USA}

\author{Changhyun Ko}
\affiliation{School of Engineering and Applied Sciences, Harvard University, Cambridge, Massachusetts 02138 USA}

\author{You Zhou}
\affiliation{Department of Chemistry and Chemical Biology, Harvard University, Cambridge, Massachusetts 02138 USA}

\author{Eric Hudson}
\affiliation{Department of Physics, Harvard University, Cambridge, Massachusetts 02138 USA}
\affiliation{Department of Physics, The Pennsylvania State University, University Park, Pennsylvania 16802 USA}

\author{Shriram Ramanathan}
\affiliation{School of Engineering and Applied Sciences, Harvard University, Cambridge, Massachusetts 02138 USA}
\affiliation{School of Materials Engineering, Purdue University, West Lafayette, Indiana 47907 USA}

\author{Jennifer E. Hoffman}
\email[Author to whom correspondence should be addressed: ]{jhoffman@physics.harvard.edu}
\affiliation{Department of Physics, Harvard University, Cambridge, Massachusetts 02138 USA}
\affiliation{Department of Physics and Astronomy, University of British Columbia, Vancouver, British Columbia V6T 1Z4 Canada}

\author{Jason D. Hoffman}
\email{jasonhoffman@fas.harvard.edu}
\thanks{This article may be downloaded for personal use only. Any other use requires prior permission of the author and AIP Publishing. This article appeared in Appl. Phys. Lett. \textbf{120}, 151602 (2022) and may be found at \url{https://doi.org/10.1063/5.0086932}.}
\affiliation{Department of Physics, Harvard University, Cambridge, Massachusetts 02138 USA}
\affiliation{Department of Physics and Astronomy, University of British Columbia, Vancouver, British Columbia V6T 1Z4 Canada}

\begin{abstract}
We present a method for nanoscale thermal imaging of insulating thin films using atomic force microscopy (AFM), and we demonstrate its utility on VO$_2$. We sweep the applied voltage $V$ to a conducting AFM tip in contact mode and measure the local current $I$ through the film. By fitting the $IV$ curves to a Poole-Frenkel conduction model at low $V$, we calculate the local temperature with spatial resolution better than 50 nm using only fundamental constants and known film properties.  Our thermometry technique enables local temperature measurement of \textit{any} insulating film dominated by the Poole-Frenkel conduction mechanism, and can be extended to insulators that display other conduction mechanisms.
\end{abstract}

\maketitle

Versatile thermal imaging methods are needed to address urgent problems in materials science,\cite{Teyssieux2007, Reparaz2014, Kim2012} biomedical applications,\cite{Kucsko2013} semiconductor device technology,\cite{Lee2007, Tessier2007, Neumann2013, Aigouy2005, Sadat2010, Menges2016, Mecklenburg2015, Halbertal2016} and future information storage devices, including phase change memories.\cite{GiraudJAP2005, GrosseJAP2014, BosseJAP2014, YalonSciRep2017} However, few existing methods can accurately map temperatures on the nanoscale. Here, we describe a scanning probe thermal imaging technique that can operate over a wide range of temperatures, length scales, environments, and insulating samples. To demonstrate the potential of our approach, we perform high spatial resolution temperature mapping in the insulating state of a voltage-biased VO$_2$ thin film.  Upon further increasing the applied voltage, we find that the tip-sample current can not only measure, but also heat the sample and trigger a thermal insulator-to-metal transition (IMT) in this material.

\begin{figure}
	\centering
	\includegraphics[width=85mm]{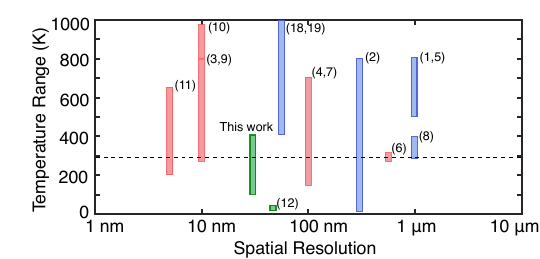}
 	\caption{Projected temperature range and demonstrated spatial resolution for thermometry techniques. The dotted line is room temperature. Pink bars represent methods that measure the temperature fluctuation about a reference temperature. Blue bars represent models where a calibration is required to measure the temperature over the applicable range. Green bars represent methods where no reference temperature or calibration is required. Techniques represented here are, from left to right: plasmon energy loss,\cite{Mecklenburg2015} Seebeck coefficient thermocouple,\cite{Sadat2010, Kim2012} thermoresistive probe,\cite{Menges2016} superconducting junction,\cite{Halbertal2016} electron thermal microscopy,\cite{BegtrupPRL2007, BrintlingerNanoLett2008} fluorescence in nanodiamonds,\cite{Kucsko2013, Neumann2013} Raman spectroscopy,\cite{Reparaz2014} infrared thermoreflectance\cite{Tessier2007} and thermography,\cite{Teyssieux2007, Lee2007} and rare earth dopant fluorescence.\cite{Aigouy2005}}
 	\label{fig:fig1}
\end{figure}

Figure \ref{fig:fig1} characterizes established thermal imaging techniques by their spatial resolution and temperature range.  Infrared thermography,\cite{Teyssieux2007, Lee2007} Raman spectroscopy,\cite{Reparaz2014} and thermoreflectance\cite{Tessier2007} are spatially limited by the detected wavelength, and typically do not achieve spatial resolution better than 100 nm.\cite{Brites2012} Fluorescence of rare earth particulates\cite{Aigouy2005} depends on the size of the particulates, and has been demonstrated down to 1 $\mu$m. Fluorescence in nanodiamonds,\cite{Kucsko2013, Neumann2013} Seebeck coefficient thermocouples,\cite{Sadat2010, Kim2012} thermoresistive probes,\cite{Menges2016} and scanning transmission electron microscopy of local plasmon mode\cite{Mecklenburg2015} techniques have achieved spatial resolutions below 100 nm, however these methods require an initial temperature calibration, and they measure only the relative temperature.  Electron thermal microscopy using nanoparticle melting\cite{BegtrupPRL2007, BrintlingerNanoLett2008} or nanowire deformation\cite{GuoNatComm2014} has also achieved spatial resolution below 100 nm, but requires calibrated finite element modeling to relate measured quantities to temperature over the full field of view.  Scanning superconducting junctions give excellent absolute sensitivity, but are applicable only in a narrow temperature range.\cite{Halbertal2016}

\begin{figure}
	\centering
	\includegraphics[width=86mm]{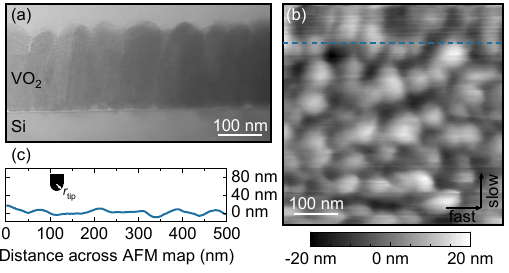}
 	\caption
 	{
 	    \label{fig:fig2}
        (a) Cross-sectional transmission electron microscopy (TEM) image of the VO$_2$ film, showing an average thickness of 187 nm and average grain diameter of 70 nm.
        (b) AFM image of the surface of the VO$_2$ film, whose varying height is used together with (a) to determine the local film thickness.
        (c) Trace along the blue dashed line of the AFM topography, using the same scale as (a), showing that the grain size is in agreement with the TEM image, as well as comparison to the estimated tip radius $r_\text{tip} \approx$ 15 nm.
    }
\end{figure}

Here, we demonstrate an approach to quantify the local temperature of an insulating thin film based on the temperature dependence of the Poole-Frenkel (PF) conduction mechanism.\cite{Frenkel1938}  In PF conduction (Eqns.\ \ref{eqn:PFeq} and \ref{eqn:PFlin}), the current depends on the dielectric constant, local film thickness, and local temperature.  We use conductive atomic force microscopy (CAFM) operating in contact mode to measure a two-dimensional grid of surface topography and simultaneous current-voltage ($IV$) curves on a VO$_2$ film. From the average film thickness measured by transmission electron microscopy (TEM) (Fig.\ \ref{fig:fig2}a) and the local topographic variation (Fig.\ \ref{fig:fig2}b), we determine the local film thickness with spatial resolution limited only by the AFM tip geometry.  We use the known bulk temperature dependence of the film dielectric constant $\varepsilon(T)$ to fit each individual $IV$ curve to the PF mechanism. The local temperature is thus the only free parameter, so we can perform quantitative temperature imaging without requiring a reference point or additional calibration.  Like existing thermal measurement techniques,\cite{Teyssieux2007, Reparaz2014, Kim2012, Kucsko2013, Lee2007, Tessier2007, Neumann2013, Aigouy2005, Sadat2010, Menges2016, Mecklenburg2015, Halbertal2016, GrosseJAP2014, BosseJAP2014, YalonSciRep2017, GiraudJAP2005}  which necessarily exchange heat with the sample and perturb its temperature, however minutely, our CAFM-based thermometry heats the sample through Joule dissipation.

We chose VO$_2$ as a model system to establish the utility of our technique, due to its widespread interest\cite{LiuMatToday2018} and technological potential.\cite{Pergament2013}  VO$_2$ undergoes an IMT where the resistivity decreases by up to five orders of magnitude as the temperature is increased through 341 K.\cite{Morin1959} Recently, voltage-triggered switching has also been demonstrated in VO$_2$,\cite{Qazilbash2008, Stefanovich2000, Kim2004, Ko2008a, RuzmetovJAP2009, Kim2010, Wu2011} but the underlying mechanism remains controversial. Previous CAFM studies suggest the importance of Joule heating,\cite{Kim2010} while simulations matched to experimental data find the IMT in VO$_2$ to be electrothermally driven.\cite{ValmianskiPRB2018} Still other simulations indicate that the current through the insulating state is insufficient to heat most 2-terminal device geometries to the bulk transition temperature.\cite{Gopalakrishnan2009}  Here, we use CAFM-based thermometry to acquire a nanoscale map of the lower bound of the local temperature immediately preceding the resistive phase change under applied voltage bias.  We find an average local film temperature of at least $337\pm4$ K, which suggests that in this geometry and dc measurement mode, the applied voltage results in significant Joule heating that contributes to triggering the IMT.

The VO$_2$ film was grown by radio frequency sputtering from a VO$_2$ target on a heavily As-doped Si(001) substrate  with $\rho$ = 0.002 -- 0.005 $\Omega$ cm.\cite{Kim2010}  Room temperature x-ray diffraction measurements\cite{Kim2010} are consistent with a polycrystalline, monoclinic VO$_2$ phase.  The cross-section TEM image in Fig.\ \ref{fig:fig2}a shows an average film thickness of 187 nm, with a root-mean-square roughness of 6.3 nm, in agreement with the AFM topography in Figs.\ \ref{fig:fig2}(b, c).

\begin{figure}
	\centering
	\includegraphics[width=86mm]{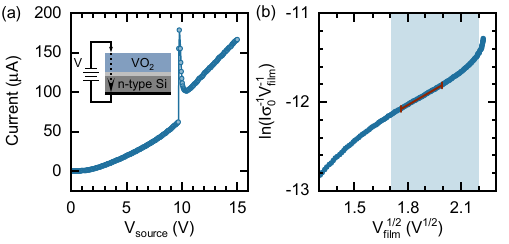}
 	\caption
 	{
        \label{fig:fig3}
        (a) Typical current versus voltage ($IV$) curve, raw data.  The sharp increase in current indicates the IMT, but the subsequent spike arises from a stray capacitance in the substrate and external circuit.  Inset: schematic of the data collection geometry.
 	    (b) The PF-linearized data shows
        $\ln(I\sigma_0^{-1} V_\text{film}^{-1}) \propto V^{1/2}$ up to the transition. The constant of proportionality $\sigma_0$, depends on the initial conductivity of the film and has been set to 1 S for the purpose of the plot.  This constant changes the offset, but not the slope of the PF-linearized data, and thus does not affect the temperature measurement. The red line indicates the linear fit used to extract the temperature through the PF model, while the shaded blue region indicates the region where Poole-Frenkel conduction may be applicable. The contact resistance and external circuit resistance have been removed in (b), leaving only the voltage through the VO$_2$ film.
     }
\end{figure}

All measurements are carried out in high vacuum at room temperature using a home-built AFM with a Au-coated cantilever tip in contact mode.  The cantilever deflection is measured by an interferometer with a wavelength of 1550 nm.  A feedback loop controls the $z$ position of the sample to hold the interferometer signal constant, thus maintaining a constant contact force of around 280 nN between the cantilever (with typical spring constant $k_c$ = 40 N/m)\cite{Cantilever} and sample.\cite{Kim2010}  We use the manufacturer's nominal tip radius along with feature sizes in the trace in Fig.\ \ref{fig:fig2}c to estimate a tip contact radius of 15 nm.  It has been shown that applying stress along the \textit{c}-axis shifts the IMT of a VO$_2$ film at a rate of $-1.2$ K/kbar.\cite{Ladd1969}  Using the tip contact radius of 15 nm, this force corresponds to a pressure of 3.9 kbar, which lowers the transition temperature by at most 4.7 K. The $z$ position is used to map the local topography in a 500 nm $\times$ 500 nm region as shown in Fig.\ \ref{fig:fig2}b.  We simultaneously gather local electronic information in this region by sweeping the voltage and measuring the current at each point in a grid pattern with approximately 2 nm separation.  We cycle the sample bias from 0 V to 15 V and back, measuring the current at 0.05 V intervals.  A typical $IV$ curve is shown in Fig.\ \ref{fig:fig3}a.  Four consecutive $IV$ sweeps are performed at each point on the map, and are found to be consistent with each other, ruling out sudden changes in contact resistance or film quality.  All data presented here were obtained from the second sweep. This mapping method should be distinguished from previous work on VO$_2$, which presented only individual $IV$ curves at a single location of the sample \cite{Kim2010,Wu2011} or current maps acquired by scanning across the sample with a constant sample bias.\cite{Kim2010}

At low voltage bias, the VO$_2$ film is in an insulating state.  As the bias is increased, we observe a sharp jump in current, indicating a transition to the metallic state.  Due to series resistance from a combination of the tip-sample contact, the spatially-dependent interface layer observed in the TEM image in Fig.\ \ref{fig:fig2}(a), and the external CAFM circuitry, the measured resistance in the metallic state appears three orders of magnitude larger than expected for the VO$_2$ film alone.  Furthermore, stray capacitance from the same external elements leads to large current transient immediately following the transition.  The excess resistance and current jump are not a general feature of CAFM spectroscopy,\cite{SchaadtJVSTB2004, Kim2010} but rather are unique to our setup and may be mimimized through future improvements, such as the use of shielded coplanar probes.\cite{RommelJVSTB2013}  As we are interested only in the voltage through the VO$_2$ film in the insulating state, we subtract off the series voltage drop through the substrate and external circuitry (Supplementary Information, section 1).

\begin{figure*}
	\centering
	\includegraphics[width=132mm]{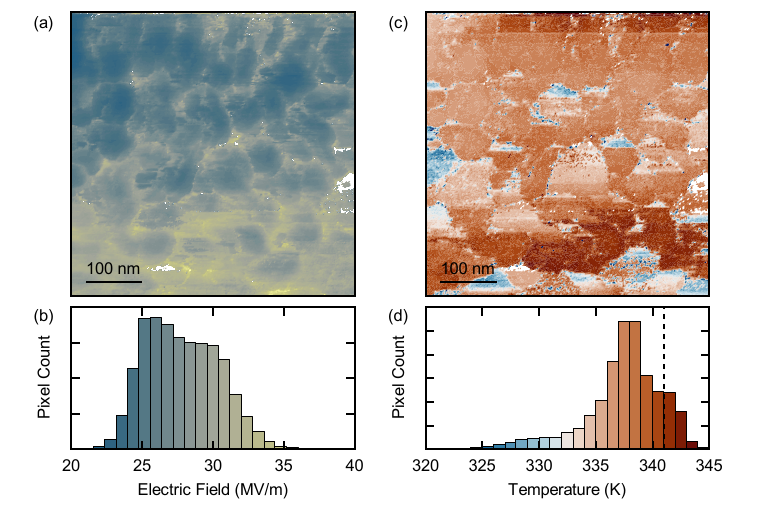}
 	\caption{Simultaneous measurement of the local electric field and local temperature of the insulator to metal transition (IMT) of VO$_2$.  (a) Map of the electric field at the IMT. (b) Histogram of the map from (a). (c) Map of the local temperature immediately preceding the IMT. All maps are from the same region of the sample shown in Fig.\ \ref{fig:fig2}b. (d) Histogram of the temperature map in (c). The dotted line denotes the bulk IMT temperature of VO$_2$. We note the sharp decrease of values greater than the bulk IMT temperature of 341 K.}
 	\label{fig:fig4}
\end{figure*}

For each $IV$ curve on the grid, we determine the voltage across the film that is required to trigger the IMT.  The local film thickness $d$, determined from the average thickness measured through TEM (Fig.\ \ref{fig:fig2}a) in combination with the local topography (Fig.\ \ref{fig:fig2}b), is used to compute the local electric field at the transition.  Figure \ref{fig:fig4}a shows a map of the electric field at the IMT, with the corresponding histogram shown in Fig.\ \ref{fig:fig4}b.  We find the average field across the VO$_2$ film at the IMT is 28 MV/m, with a standard deviation of 3 MV/m, consistent with previous reports of an electric field-driven phase transition in VO$_2$.\cite{RuzmetovJAP2009, Wu2011, Ko2008a} In the polycrystalline film studied here, we use the tip-limited spatial resolution to show that there is little variation in electric field at the transition within each individual VO$_2$ grain.  However, the grain edges display a higher electric field at the transition, which may be due to local variations in strain or stoichiometry at the boundary,\cite{Kim2010} or the AFM tip contacting multiple grains at once.

To quantify the local temperature, we investigate the shape of the $IV$ curve immediately preceding the transition. Previous experiments have shown that VO$_2$ is dominated by PF conduction in the temperature and electric field regimes investigated here\cite{Ko2008a, RuzmetovJAP2009, YangPRB2010} (see additional references S13--S22 confirming PF conduction in VO$_2$ in the Supplementary Information).  In agreement with these previous studies, we find that PF conduction is the dominant mechanism in our VO$_2$ film.  While PF conduction has previously been \textit{observed} in VO$_2$ films,\cite{Ko2008a, RuzmetovJAP2009, Wu2011, YangPRB2010} here we \textit{utilize} it to quantify the local temperature; this is the primary advancement of our current work.

In an insulating system dominated by PF conduction, the combination of high temperatures and strong electric fields excite trapped electrons into the conduction band,\cite{Chiu2014} resulting in a current density $\vec{J}$ given by
\begin{equation}
	|\vec{J}| = e \mu n |\vec{E}|~\exp\left[ \frac{e(e |\vec{E}| / \pi\varepsilon_0\varepsilon)^{1/2} - e\phi_T}{k_B T}\right].
	\label{eqn:PFeq}
\end{equation}
Here, $e$ is the electron charge, $\mu$ is the electron mobility, $n$ is the density of states of the conduction band, $\vec{E}$ is the electric field, $\varepsilon_0$ is the permittivity of free space, $\varepsilon$ is the relative permittivity of the insulator, $\phi_T$  is the average trap depth, $k_B$ is the Boltzmann constant, and $T$ is the temperature of the film.  Upon rearranging Eqn.\ \ref{eqn:PFeq} for $I$ and $V$, we obtain
\begin{equation}
	\ln\frac{I}{V} = \frac{e^{3/2}}{(\pi \varepsilon_0 \varepsilon d)^{1/2}~k_BT}V^{1/2} + \ln (\sigma_0) - \frac{e\phi_T}{k_BT},
	\label{eqn:PFlin}
\end{equation}
where $\sigma_0 = ae\mu n/d$, $a$ is the contact area of the tip, and $d$ is the film thickness.  Figure \ref{fig:fig3}b shows a PF-linearized $IV$ curve, where the red line is the PF slope fit to Eqn.\ \ref{eqn:PFlin}.

We fit each PF-linearized $IV$ curve, and use the slope $P \equiv e^{3/2}/[(\pi \varepsilon_0 \varepsilon d)^{1/2}\,k_BT]$ to calculate the local temperature of the insulating state below the IMT at each point on the map.  Many factors in $P$ are fundamental constants, leaving only the film thickness and relative permittivity to be measured, before determining $T$.  The local film thickness $d$ was determined from Fig.\ \ref{fig:fig2}.  The relative permittivity was measured on a separate VO$_2$/Si(001) film, in an experiment where the global temperature of the film was varied from 290 K to 333 K (Supplementary Information, section 2).  The interpolated function $\varepsilon(T)$ is then used to self-consistently determine the temperature associated with each $IV$ curve in the map.  The calculated temperature map and corresponding histogram are shown in Figs.\ \ref{fig:fig4}(c, d).  Similar to the map of the electric field at the IMT, individual grains are discernible due to abrupt changes at grain boundaries.

We find an average temperature of 337 K prior to the IMT over the region of interest, with a standard deviation of 4 K.  We calculate an uncertainty of 4 K for each individual temperature pixel, from uncertainties in $d$, $\varepsilon(T)$, and fitted slope (Supplementary Information, section 3).  The average temperature is well above room temperature and just below the known bulk transition temperature of 341 K,\cite{Morin1959, Qazilbash2008} which may be suppressed by tip-induced pressure.\cite{Ladd1969}  This finding suggests that Joule heating has locally elevated the sample temperature very close to the transition.  The sharp decrease in our measured temperature histogram above 341 K supports the validity of our technique.  We note that the finite voltage range required for fitting means that the extracted temperature is an average over an interval shortly preceding the transition; the instantaneous temperature at the end of this interval is presumably even larger and closer to the bulk transition temperature.

We have implemented a nanoscale thermal imaging technique, and used it to address the long debate of whether or not the electronic phase transition in VO$_2$ can be triggered by an applied electric field, finding that in our geometry the applied voltage has heated the film close to its transition temperature.\cite{RuzmetovJAP2009, Wu2011, Ko2008a} However, this technique may also be applied to other materials that display PF conduction with or without phase transitions, including Pr$_2$O$_3$,\cite{Chiu2009} Ge$_2$Sb$_2$Te$_5$,\cite{GotohJNCS2008} and simple binary oxides such as ZnO,\cite{Chang2008} SnO$_x$, AlO$_x$, CeO$_x$ and WO$_x$,\cite{Lim2015} which have attracted attention as alternative gate dielectrics,\cite{Raffaella2003} phase change memories,\cite{GotohJNCS2008} and resistive switching devices.\cite{Chang2008}

Our approach can be extended to insulating materials dominated by other conduction mechanisms, such as Schottky field emission, Fowler-Nordheim tunneling, thermionic-field emission, hopping conduction, ohmic conduction, and ionic conduction.\cite{Chiu2014}  Each of these mechanisms can be linearized in terms of current and voltage, with a slope that is dependent on the film temperature and various parameters, depending on the mechanism.  These parameters include a combination of fundamental constants and film-dependent properties.  For example, in addition to fundamental constants, the hopping conduction linearized slope depends on film thickness and the hopping distance between traps, both of which could be measured.   This CAFM thermometry technique could therefore be extended to any film dominated by one of these conduction mechanisms once the film-dependent parameters are known.

In conclusion, we developed a CAFM-based technique to map local temperature in an insulating thin film, and applied it to VO$_2$.  We found that the applied voltage increased the temperature of the insulating VO$_2$ film to around 337 K through Joule heating.  This finding indicates that the IMT is not solely triggered by the applied electric field in the geometry and timescale of our experiment.  This technique is capable of measuring the local temperature of a film with spatial resolution limited only by the contact area of the tip (radius $\approx 15$ nm in our experiment, but potentially as small as 10 nm \cite{Hafner1999}) and a temperature uncertainty of 4 K, without the use of a reference temperature or unknown free parameters.  The use of $IV$ curves acquired through CAFM as a local temperature probe may be applied to a wide range of insulating thin film systems to address questions in materials science and novel nanoscale devices.

\section*{Supplementary Material} See the supplementary materials for calculations of the voltage drop across the film, measurements of the dielectric constant of VO$_2$, details of our data analysis protocols and estimates of temperature uncertainty, and a description of other conduction mechanisms in insulating materials.

\begin{acknowledgments}
The experimental work was supported by the National Science Foundation under Grant No.\ DMR-1231319 (STC Center for Integrated Quantum Materials) and the sample fabrication was supported by AFOSR Grant No.\ FA9550-08-1-0203. J.D.H.\ acknowledges support from the Gordon and Betty Moore Foundation's EPiQS Initiative through Grant No.\ GBMF4536. A.S.\ acknowledges support from the Canadian NSERC CGS-M graduate fellowship. The authors thank Harry Mickalide and Kevin O'Connor for useful discussions.
\end{acknowledgments}

The data that support the findings of this study are available from the corresponding authors upon reasonable request.

\end{document}